\documentclass[aps,prb,reprint,superscriptaddress]{revtex4-1}
\usepackage{color}
\usepackage{graphicx}
\usepackage{hyperref}

\begin{document}

\preprint{Trs/1.2}

\title{Magnetic field-induced exchange effects between $Mn$ ions and free carriers in $ZnSe$ quantum well through the intermediate nonmagnetic barrier studied by photoluminescence}

\author{D. M. Zayachuk}
 \affiliation{Lviv Polytechnic National University - 12 Bandera St, 79013
Lviv, Ukraine}
\author{T. Slobodskyy}
 \email{Taras.Slobodskyy@iss.fzk.de}
 \affiliation{Physikalisches Institut der Universit\"{a}t W\"{u}rzburg - 97074 W\"{u}rzburg, Germany}
 \affiliation{Present address: Institute for Synchrotron Radiation, Karlsruhe Institute of Technology - 76344 Eggenstein-Leopoldshafen, Germany}
\author{G. V. Astakhov}
 \affiliation{Physikalisches Institut der Universit\"{a}t W\"{u}rzburg - 97074 W\"{u}rzburg, Germany}
 \affiliation{A. F. Physico-Technical Institute, Russian Academy of Sciences - 194021 St. Petersburg, Russia}
\author{A. Slobodskyy}
 \affiliation{Light Technology Institute, Karlsruhe Institute of Technology (KIT), Kaiserstr. 12, 76131 Karlsruhe, Germany}
 \affiliation{Zentrum f\"{u}r Sonnenenergie- und Wasserstoff-Forschung Baden-W\"{u}rttemberg, Industriestr. 6, 70565 Stuttgart, Germany}
\author{C. Gould}
 \affiliation{Physikalisches Institut der Universit\"{a}t W\"{u}rzburg - 97074 W\"{u}rzburg, Germany}
\author{G. Schmidt}
 \affiliation{Physikalisches Institut der Universit\"{a}t W\"{u}rzburg - 97074 W\"{u}rzburg, Germany}
 \affiliation{Present and permanent address: Institut f\"{u}r Physik, Universit\"{a}t Halle - 06099 Halle, Germany}
\author{W. Ossau}
 \affiliation{Physikalisches Institut der Universit\"{a}t W\"{u}rzburg - 97074 W\"{u}rzburg, Germany}
\author{L. W. Molenkamp}
 \affiliation{Physikalisches Institut der Universit\"{a}t W\"{u}rzburg - 97074 W\"{u}rzburg, Germany}

\date{\today}

\begin{abstract}

Photoluminescence (PL) of the
50~nm~$Zn_{0.9}Be_{0.05}Mn_{0.05}Se$/
$d$~nm~$Zn_{0.943}Be_{0.057}Se$/ 2.5~nm~$ZnSe$/ 30~nm~$Zn_{0.943}Be_{0.057}Se$ structures is investigated as a function of magnetic field
($B$) and thickness ($d$) of intermediate $Zn_{0.943}Be_{0.057}Se$ nonmagnetic barrier between the
$Zn_{0.9}Be_{0.05}Mn_{0.05}Se$ semimagnetic barrier and $ZnSe$ quantum well
at the temperature 1.2~K. The rate of the shift of different PL bands
of the structures under study is estimated in low and high magnetic
fields. The causes of the shift rate increase under pass from low to
high magnetic fields are interpreted.  The peculiarities of the effect
of the intermediate barrier on the luminescence properties of the
structures are presented. It is shown that deformation of adjacent
layers by the barrier plays a crucial role in the formation of these
properties, especially in forming the $Mn$ complexes in the
$Zn_{0.9}Be_{0.05}Mn_{0.05}Se$ layer. 
The change of the band gap as well as of the donor and acceptor levels energies 
under the effect of biaxial compression of the
$Zn_{0.9}Be_{0.05}Mn_{0.05}Se$ layer by
the $Zn_{0.943}Be_{0.057}Se$ are
estimated. It is concluded that the
$Zn_{0.943}Be_{0.057}Se$ intermediate
barrier also appreciably changes the effect of giant Zeeman splitting
of the semimagnetic $Zn_{0.9}Be_{0.05}Mn_{0.05}Se$ barrier
energy levels on the movement of the energy levels of
$ZnSe$ quantum well in a magnetic field and
on polarization of the quantum well exciton emission.

\end{abstract}

\pacs{78.55.Et, 73.21.Fg, 78.20.Ls}


\maketitle

\section{INTRODUCTION}

Quantum heterostructures containing layers of diluted magnetic
semiconductors (DMS), usually a $3d$
$Mn$ or $Fe$ based transition metal, have been
extensively studied in the literature. The main focus has been made on
both fundamental and practical applications, especially these designed
for different spin-electronic devices. Presence of transition element
ions provides the conditions for magnetic tuning of the
heterojunction band alignment due to extraordinarily large
spin-splitting of the DMS bands due to the exchange
interaction between the $s$ and $p$ band electrons and
$3{d^5}$ electrons associated with the
$3d$ element ions \cite{ref1, ref2, ref3}. The $s$, $p -
d$ exchange interaction between the local moments and the band
electrons gives rise to a rich spectrum of collective magnetic
behavior. When an external magnetic field is applied to DMS, they
exhibit two distinct band gaps, one for each spin direction \cite{ref1}. This splitting of conduction and valence bands (giant Zeeman splitting) can lead to a sizable spin polarization of the carriers in
the DMS. This property is used to inject spin-polarized carriers into
nonmagnetic semiconductors \cite{ref4, ref5}.

Two typical approaches are usually applied to the fabrication of DMS
heterostructures:  a heterostructure with a DMS layer as a quantum well
(QW) \cite{ref6} or a heterostructure with DMS layers as quantum barriers \cite{ref7, ref8}. In
the first case, both the free carriers and the
$3d$ element ions are located in the same
layer of a QW. This leads to a strong interaction between the free
carriers and the localized
$3d$-electrons of ions which
intensifies the effects caused by the magnetic field. However, the
presence of magnetic impurities in a quantum well stimulates spin
relaxation processes \cite{ref9, ref10, ref11}. Hence, it is preferable to
separate the carriers from the magnetic media \cite{ref12, ref13, ref14}. In
this case, the exchange interaction between the free
$2D$ carriers of nonmagnetic quantum well
and the ions of magnetic impurities in the barrier is driven by
penetration of the carrier wave function tails into the barrier.

The effective depth of a well, represented by the difference between
the positions of the band edges in adjacent layers, is strongly
dependent on the magnetic field. At the same time, the energy of
quantization levels in a shallow well strongly depends on the well
depth. Therefore, the electron states in a nonmagnetic quantum well
acquire some characteristics of the DMS materials, specifically
sensitivity to the magnetic field \cite{ref1}.

It seems interesting to investigate the effect of magnetic field on the electron states in a nonmagnetic quantum well effected by a semimagnetic barrier. The barrier changes its height and exchange interaction with the states. Separation of the two effects can provide an additional degree of freedom in fabricating spintronics devices. In particular, an additional nonmagnetic layer introduced between a semimagnetic barrier and a nonmagnetic quantum well may be used for this purpose. Since, in the range of our experimental conditions, the nonmagnetic barrier height only negligibly depends on a magnetic field,
the changes of an exchange interaction contribution will dominate in
the field behavior of the energy characteristics of the QW. In this
paper we show such a possibility using the example of quantum
structures based on the $ZnSe$ quantum well with the
$Zn_{0.9}Be_{0.05}Mn_{0.05}Se$ and
$Zn_{0.943}Be_{0.057}Se$ barriers. The
barrier compositions were selected based on the condition of an
approximate parity of the band gap of the barriers at zero magnetic
field. In order to study the field-induced exchange effects through
the intermediate nonmagnetic barrier between
$Mn$ ions and free carriers in
$ZnSe$ quantum well, $Zn_{0.943}Be_{0.057}Se$ barrier of
different thickness was introduced between the $ZnSe$ and
$Zn_{0.9}Be_{0.05}Mn_{0.05}Se$ layers. Photoluminescence (PL) properties of the structure in magnetic
fields up to 5.25~T were investigated.

\section{EXPERIMENTAL DETAILS}

\begin{figure}\
 \centerline{\includegraphics [width=7cm]{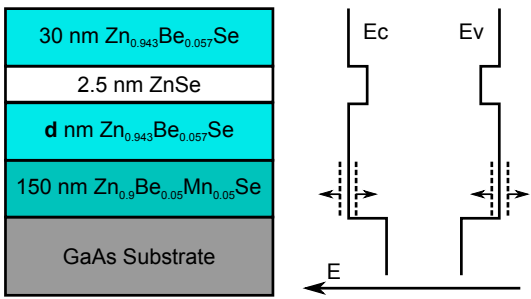}}
 \caption{Schematic view of the sample layout (left) with the
corresponding energy band profile (right). The spin-split sub-bands of
the conduction band (${E_C}$) and the
valence band (${E_V}$) in the magnetic field
are depicted by arrows. The
$Zn_{0.943}Be_{0.057}Se$ barrier
thickness $d$ was varied as 0 (named
$a1$), 2.5
($b1$), 7.5
($c1$), and 12.5~nm
($d1$).}
 \label{1}
\end{figure}

The samples used in the PL experiments were grown by molecular beam
epitaxy on a $GaAs$ substrate. The sample layout is depicted in
Fig.~\ref{1}. The nonmagnetic space layer $d$ is
used for the purpose of varying the magnetic interaction in the
structure.

PL spectra at 1.2~K in magnetic fields up to 5.25~T at Faraday
geometry were measured to study the peculiarities of the exchange
interaction between the $Mn$ ions
and free carriers. For optical excitation, we used a stilbene-3 dye
laser pumped by ultraviolet lines of an Ar-ion laser. For non-resonant
excitation, the laser energy is tuned to $E_{exc}$ = 2.94 eV
exceeding the band gap of the
$Zn_{0.9}Be_{0.05}Mn_{0.05}Se$
barrier.

\section{THE EXPERIMENTAL PL SPECTRA}

\begin{figure}\
 \centerline{\includegraphics [width=7cm]{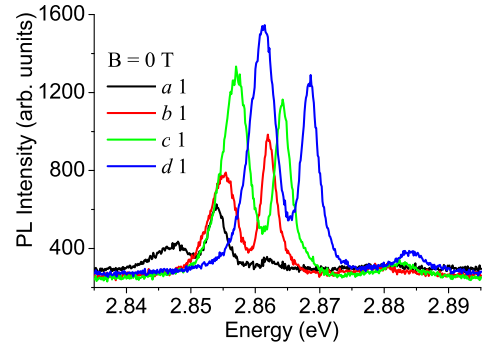}}
 \caption{The PL spectra of the 150~nm~$Zn_{0.9}Be_{0.05}Mn_{0.05}Se$~/ $d$~nm~$Zn_{0.943}Be_{0.057}Se$~/ 2.5~nm~$ZnSe$~/ 30~nm~$Zn_{0.943}Be_{0.057}Se$ structures in zero magnetic fields: T~= 1.2~K. $\hbar {\nu _{ex}}$ = 2.94 eV.}
 \label{2}
\end{figure}

Fig.~\ref{2} shows the PL spectra for the structures under study without the
application of magnetic field. One can see that the shape of PL
spectrum critically depends on the presence of an intermediate
$Zn_{0.943}Be_{0.057}Se$ barrier between
the $ZnSe$ QW and the semimagnetic
$Zn_{0.9}Be_{0.05}Mn_{0.05}Se$ layer.
This concerns the two "evident" spectra characteristics, i. e., the
number of the PL bands and their intensity. The most striking feature
of this dependence is a decrease of the number of PL bands by one for
the structures with an intermediate barrier.

\begin{figure*}\
 \centerline{\includegraphics [width=13cm]{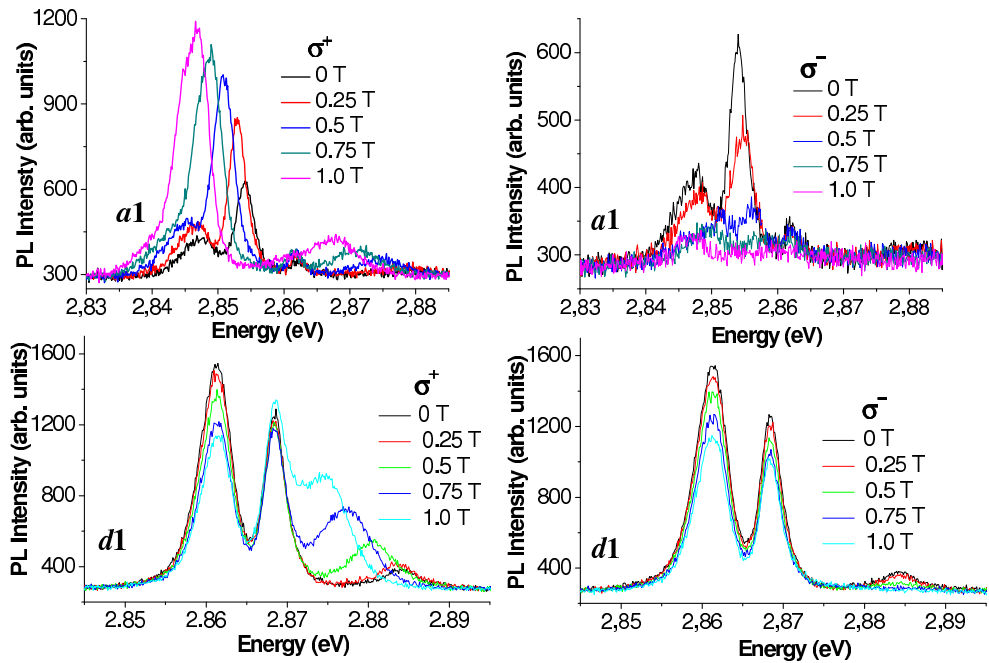}}
 \caption{The PL spectra of ${\sigma ^ + }$
and ${\sigma ^ - }$-polarization of the $a1$(top) and $d1$(bottom) structures in magnetic fields.}
 \label{3}
\end{figure*}

The composition of the structure also affects the behavior of the
PL bands in the magnetic field where the bands split into two
components with ${\sigma ^ + }$ and ${\sigma ^ - }$-polarization. Fig.~\ref{3} shows
this effect using the example of different polarization PL spectra in a magnetic field up to 1~T for $a1$ and $d1$ structures.

\begin{figure*}\
 \centerline{\includegraphics [width=13cm]{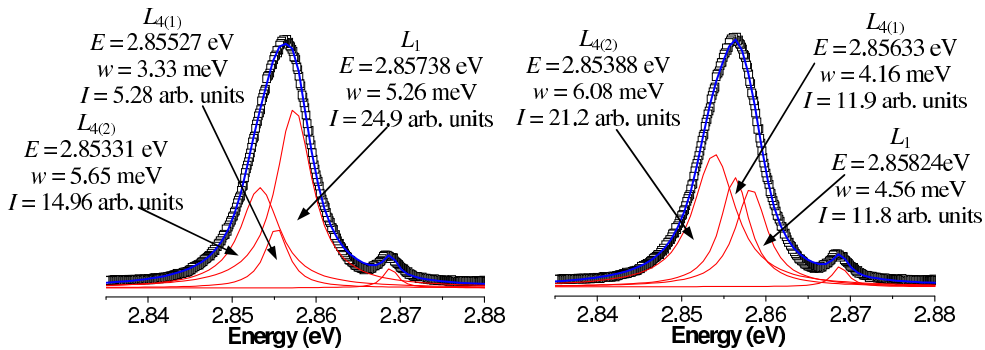}}
 \caption{Two examples of a multitude of decompositions of the
experimental ${\sigma ^ + }$-polarized PL
spectrum of the $d1$ structure in 5~T
magnetic fields on the Lorentz components satisfying a requirement of a
domination of the intensity of the heavy excitons (the left band) over
the intensity of the light excitons (the second band from the left) in
the $ZnSe$ QW. Classification of the
${L_1}$ -
${L_4}$ bands see below in Sec.~\ref{EnergyPL}, their
nature - in Sec.~\ref{discussion}.}
 \label{4}
\end{figure*}

In order to quantitatively analyze the structure composition effect on
the behavior of PL spectra in a magnetic field, it is necessary to
separate the spectra into elementary components. This turns out to be very
complicated problem, especially in the range of high magnetic fields
where different PL bands superimpose one above another. As a result, the experimental PL spectrum can be reproduced in different ways by the
same number of elementary bands. As an example, the data in Fig.~\ref{4}
illustrate this.

In view of the mentioned ambiguity, decomposition of the experimental spectra into elementary constituents was carried out using the method of averaging. Details of the method of decomposition will be discussed more in detail elsewhere. Here we emphasize that it is well known \cite{ref12} that for the zinc-blende structure, the transition matrix element of
the transitions involving heavy holes (hh) is three times larger than
that involving light holes (lh). Therefore, only those decompositions
were taken into account where the PL intensity of the heavy excitons
was larger than the PL intensity of the light excitons. It will be
shown below that exactly these excitons form the longest wave PL bands
of the structures under study. Not less than 20 different
decompositions covering the maximum range of parameter variations for
each spectrum were used to determine the averaged values of the
parameters. An analysis showed that the deviation of experimental
values of the parameters of different decompositions from their
averaged values in any structures do not exceed: (i)
$\pm{}$~0.0005~eV for the energy of a short wave
component of the spectra and $\pm{}$~0.0015~eV for
the long wave components; (ii) $\pm{}$~20~\% of the
averaged value for the full width at half maximum (FWHM) of the bands;
$\pm{}$~60~\% of the averaged value for the
intensity of the bands. The total intensity of PL is reproduced by the
intensity of the Lorentz components within the error of no more than
10~\% of the true value.

\section{ENERGY POSITION OF THE PL BANDS}
\label{EnergyPL}

\begin{figure*}\
 \centerline{\includegraphics [width=13cm]{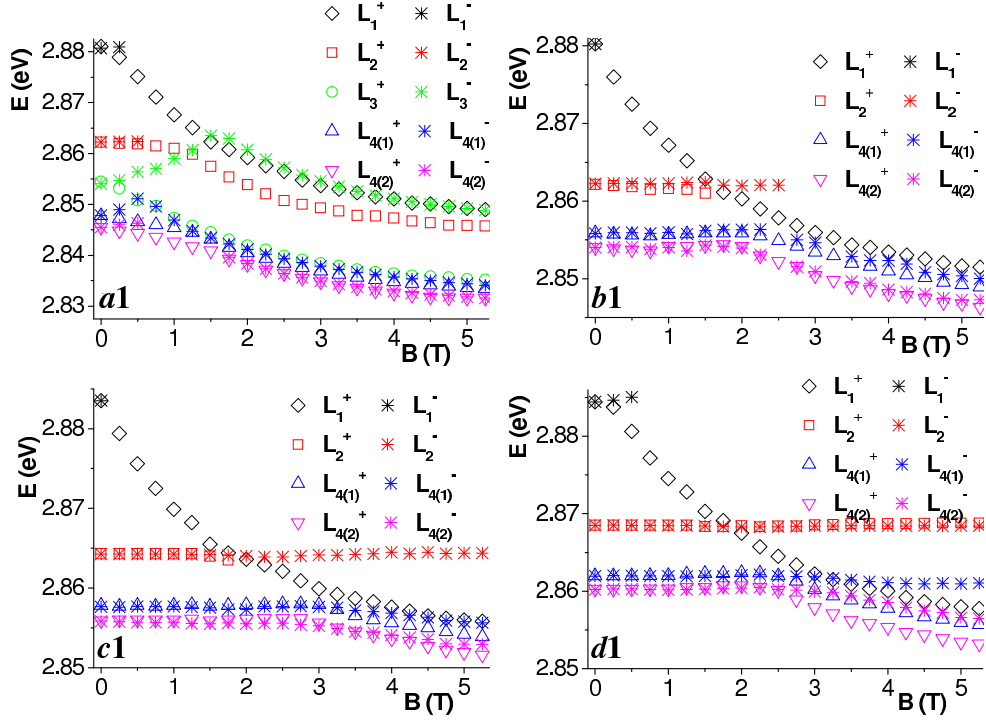}}
 \caption{Magnetic field dependence of the energy positions of the
Lorentz components of both ${\sigma ^ + }$
($L_1^ + $ -
$L_{4(2)}^ + $) and
${\sigma ^ - }$-polarizations
($L_1^ - $ -
$L_{4(2)}^ - $).}
 \label{5}
\end{figure*}

Fig.~\ref{5} shows the magnetic field dependences on the energy positions of different PL bands for the structures under study. Let us examine these dependences from viewpoint of general and distinctive features in different structures.

\textit{The ${L_1}$ band}: It is
the highest energy PL band in zero magnetic fields.  Its behavior in a
magnetic field is the same for all structures. The
${\sigma ^ + }$-polarized
$L_1^ + $ band shifts to the long-wave
range, if $B$ increases. The
${\sigma ^ - }$-polarized
$L_1^ - $ band tends to shift in the
opposite direction in a magnetic field and practically disappears when
$B$~$>$~0.5~T.

\textit{The ${L_2}$ band}: It is
the next PL band in zero magnetic fields. It behaves differently in different structures. ${\sigma ^ +}$\textit{-polarization}: (i) the $a1$ structure: the $L_2^+ $ band slightly changes its position in low ($<$ 1~T) magnetic fields and shifts appreciably to the long-wave range in high magnetic fields, when $B$ increases; (ii) the
$b1$ and $c1$
structures: the $L_2^ + $ band energy
practically does not depend on the magnetic field for
$B$ $<$~1.25~T, then
starts to decrease somewhat. The band disappears in the magnetic field
where its energy becomes equal to the $L_1^ +$ band energy; (iii) the $d1$
structure: the $L_2^ + $ band position does
not depend on the magnetic field. ${\sigma ^ -}$\textit{-polarization}: the $L_2^ -$ band energy does not depend on the magnetic field. However,
the range of magnetic fields where the band is observed is different
for different structures. It is the narrowest
($B\leq{}$0.5~T) for the
$a1$ structure and includes all magnetic
fields under investigation for the $c1$ and
$d1$ structures.

\textit{The ${L_3}$ band}: This
band displays the strongest dependence on the structure composition. It
is observed only for the $a1$ structure
without the intermediate
$Zn_{0.943}Be_{0.057}Se$ barrier. The
$L_3^ + $ band energy decreases if
$B$ increases. On the contrary, in low
magnetic fields of approximately $B$
$<$~1.5~T, the $L_3^ - $
band energy increases if $B$ increases. If
$B$ $>$~1.5~T, the
$L_3^ - $ band shifts in the opposite
direction. In the fields above 2~T its energy position coincides with
the energy position of the $L_1^ + $ band.

\textit{The ${L_4}$ band}: This is
the smallest energy PL band in zero magnetic fields. This band is a
superposition of two components with close energy. We mark them
$L_{4(1)}^ + $, $L_{4(2)}^
+ $ and $L_{4(1)}^ - $,
$L_{4(2)}^ - $ for
${\sigma ^ + }$ and
${\sigma ^ - }$-polarization, respectively.
Their behavior is absolutely different for the structures with and
without an intermediate nonmagnetic
$Zn_{0.943}Be_{0.057}Se$ barrier.
\textit{The $a1$ structure}: (i)
the energy position of the ${L_4}$ bands
changes if $B$ changes within the whole
range of magnetic field under study; (ii) the
$L_{4(1)}^ + $ and
$L_{4(2)}^ + $ bands shift to the long-wave
range if $B$ increases; (iii) if the
magnetic field is applied, the $L_{4(1)}^ -
$ and $L_{4(2)}^ - $ bands shift to
the short-wave range but they change the direction of their shift if
$B$ $>$~0.5~T. In this
range they follow the positions of the respective PL bands of
${\sigma ^ + }$-polarization. \textit{The $b1$, $c1$, and $d1$ structures}: (i)
in low magnetic field, the energy of the bands practically does not
depend on the magnetic field and thus the $L_{4(1)}^ + $ and $L_{4(1)}^ - $ as well as
$L_{4(2)}^ + $ and
$L_{4(2)}^ - $ coincide with each other;
(ii) in high magnetic field, all bands shift to the long-wave range and
the bands with different polarizations gradually diverge. The thicker is
the intermediate barrier the larger is the divergence of both
$L_4^ + $ and $L_4^ -$ bands.

\section{FULL WIDTH AT HALF MAXIMUM OF THE PL BANDS}

\begin{figure*}\
 \centerline{\includegraphics [width=13cm]{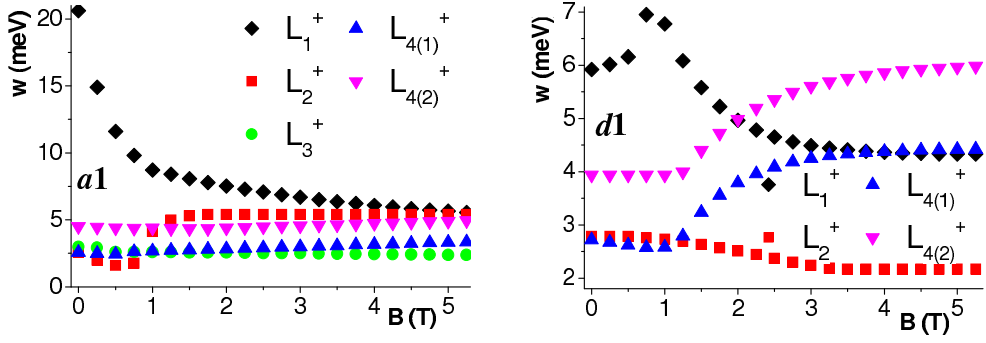}}
 \caption{Magnetic field dependence of the FWHM of the
${\sigma ^ + }$-polarized bands of the
$a1$ and $d1$
structures.}
 \label{6}
\end{figure*}

\begin{figure*}\
 \centerline{\includegraphics [width=13cm]{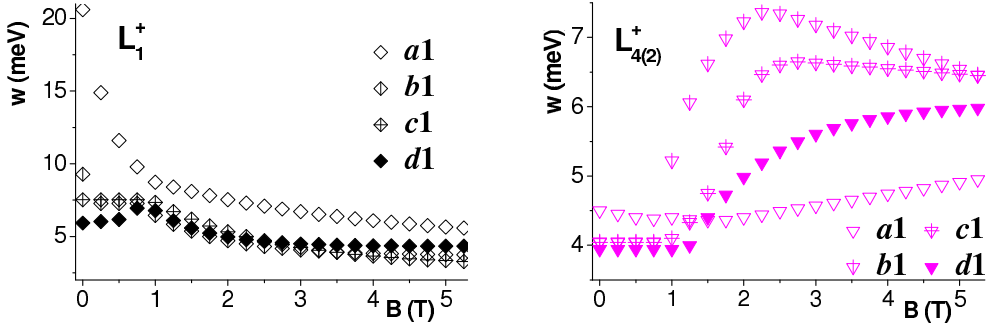}}
 \caption{Magnetic field dependence of the FWHM of the
$L_1^ + $ and $L_{4(2)}^ +
$ bands for all structures.}
 \label{7}
\end{figure*}

Fig.~\ref{6} and Fig.~\ref{7} show the effect of both the intermediate barrier and
the magnetic field on the PL bands FWHM $w$ for the
structures under study.

The main features of this effect are as follows.

\textit{The ${L_1}$ band}: FWHM of
the $L_1^ + $ band is the widest in the
$a1$ structure. In zero magnetic fields,
$\omega $ decreases if the intermediate
nonmagnetic barrier is applied and its thickness increases. In low
magnetic field $B$
$<$~1~T, the band behavior is different for
different structures: from continuous decrease in the
$a1$ structure to some increase in the
$d1$ structure if
$B$ increases. For
$B$ $>$~1~T,
$w$ monotonously decreases if
$B$ increases. Herein the band is
practically of the same width in all structures with the intermediate
$Zn_{0.943}Be_{0.057}Se$ nonmagnetic
barrier.

\textit{The ${L_2}$ band}: FWHM of
the $L_2^ + $ band increases approximately
three times under transition from low to high magnetic fields for the
$a1$ structure and somewhat decreases for
the $d1$ structure. FWHM of the
$L_2^ - $ band practically does not depend
on $B$ for the
$c1$ and $d1$
structures where it is observed for any investigated magnetic fields.

\textit{The ${L_3}$ band}: If a
magnetic field is applied, FWHM of the $L_3^ +$ band somewhat decreases and that of the
$L_3^ - $ band increases. After changing the
shift direction in a magnetic field, the $L_3^ -$ band FWHM decreases if $B$
increases. In this range of the magnetic field it is approximately four
times larger than the $L_3^ + $ band FWHM.

\textit{The ${L_4}$ band}: FWHM of the $L_{4(1)}^ + $ and $L_{4(2)}^ + $ bands practically does not depend on $B$ in low magnetic fields $B$ $<$~1~T and decreases if the intermediate nonmagnetic barrier is applied and its thickness increases. In the intermediate magnetic fields, $w$($L_{4(2)}^ +$) and $w$($L_{4(1)}^ +$)  sharply rise in the structures with an intermediate barrier. The band widening depends on the barrier thickness $d$: it is the largest for the $b1$ structure where $d$ is the smallest and vice versa - it is the smallest for the $d1$ structure where $d$ is the largest. If $B$ increases further, $w$($L_{4(2)}^ + $) and $w$($L_{4(1)}^ + $) starts to decrease. The FWHM decrease slackens if the barrier thickness increases. For the $d1$ structure, $w$($L_{4(2)}^ + $) and $w$($L_{4(1)}^ + $) practically do not depend on $B$ in the high magnetic field range. In this range of the field for the $a1$ structure, $w$($L_{4(2)}^ + $) and $w$($L_{4(1)}^ + $) somewhat increase if $B$ increases.

FWHM of both $L_{4(1)}^ - $ and $L_{4(2)}^ - $ bands behaves like to $L_{4(1)}^ + $ and $L_{4(2)}^ + $ bands FWHM.

\section{INTENSITY OF THE PL BANDS}

\begin{figure}\
 \centerline{\includegraphics [width=6cm]{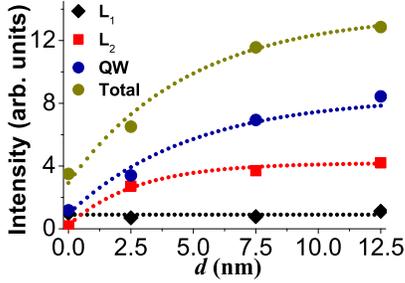}}
 \caption{Barrier thickness
$d$($Zn_{0.943}Be_{0.057}Se$)
dependence of PL signal intensity at zero magnetic field.}
 \label{8}
\end{figure}

\begin{figure*}\
 \centerline{\includegraphics [width=13cm]{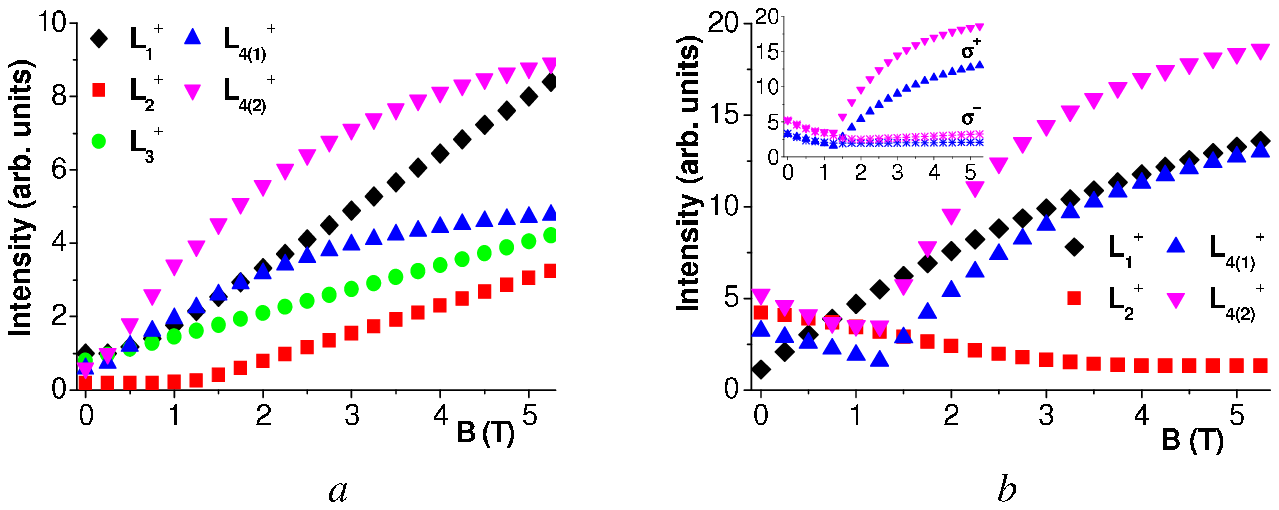}}
 \caption{Magnetic field dependence of the PL intensity of the ${\sigma ^ + }$- polarized bands for both $a1$ (left) and $d1$ (right) structures. Inset: PL intensity of both $L_{4(1)}^{}$ and $L_{4(2)}^{}$ bands of ${\sigma ^ + }$ and ${\sigma ^ - }$ polarization.}
 \label{9}
\end{figure*}

Fig.~\ref{8} and \ref{9} show an effect of both the intermediate barrier and
magnetic field on the PL band intensity $I$ for the
structures under study.

The application of the intermediate nonmagnetic barrier
$Zn_{0.943}Be_{0.057}Se$ between the
$ZnSe$ and
$Zn_{0.9}Be_{0.05}Mn_{0.05}Se$ layers
of the structure has no effect on the
${L_1}$ band intensity in zero magnetic
fields but appreciably augments the intensity of the other PL bands
- ${L_2}$, QW
$(I({L_{4(1)}}) + I({L_{4(2)}}))$ as well as
the total PL intensity. All these intensities increase if
$d$($Zn_{0.943}Be_{0.057}Se$)
increases (Fig.~\ref{8}).

Application of a magnetic field increases the intensity of the short wave $L_1^ + $ PL band in any structures. At the same time, the field effect on the behavior of the other PL bands is not that simple. For example, in a low magnetic field $B$ $<$~1~T, an intensity of the long wave $L_{4(1)}^ + $ and $L_{4(2)}^ + $ bands increase when $B$ increases in the $a1$ structure but decreases in another three structures. An intensity of the $L_2^ +$ band increases under transition from low to high magnetic fields in the $a1$ structure but decreases and comes off plateau in the $d1$ structure (Fig.~{9}).

\begin{figure*}\
 \centerline{\includegraphics [width=13cm]{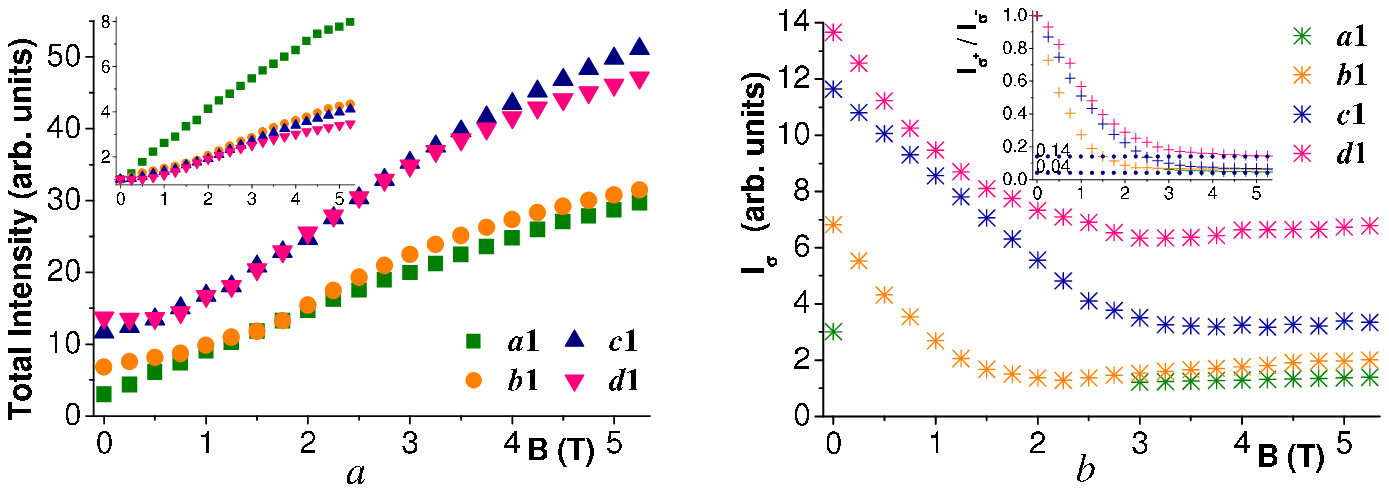}}
 \caption{Magnetic field dependence of a total intensity of ${\sigma ^ + }$ (left) and ${\sigma ^ - }$-polarized (right) PL of the structures under study. Insets: (left) - a total intensity normalized by its value in zero magnetic fields; (right) - a relation between a total intensity of ${\sigma ^ +}$ and ${\sigma ^ - }$-polarized PL.}
 \label{10}
\end{figure*}

Application of a magnetic field also increases the total intensity
${I_{\sigma  + }}$ of the
${\sigma ^ + }$-polarized emission of the
structures but decreases the total intensity
${I_{\sigma  - }}$ of the
${\sigma ^ - }$-polarized emission
(Fig.~{10}). In high magnetic fields, ${I_{\sigma  -}}$ first stabilizes and then displays a weak tendency to
increase. Herein ${I_{\sigma  - }}$ becomes
much smaller than ${I_{\sigma  + }}$. For
the top magnetic field, $B$ =~5.25~T ratio
${I_{\sigma  + }}/{I_{\sigma  - }}$ changes
from 0.04 in the $a1$ structure to 0.14 for
the $d1$ structure (see inset in Fig.~\ref{10},
($b$)).

In low magnetic fields, the character of ${I_{\sigma + }}(B)$ appreciably depends on the intermediate barrier
thickness: the thicker is the barrier the weaker is the dependence. In
high magnetic fields, an effect of the intermediate barrier thickness
on the total intensity ${I_{\sigma  + }}$ of
the structures has got a pronounced non-monotonous character. The
thinnest barrier (the $b1$ structure) does
not increase ${I_{\sigma  + }}$ with respect
to the $a1$ structure. Similar
increase of the intermediate barrier thickness from 7.5 to 12.5~nm (the
$c1$ and $d1$
structures) practically does not change the parameter
${I_{\sigma  + }}$ as well. Thus it is
possible to reach the principal increase of the total PL intensity
${I_{\sigma  + }}$ in the high magnetic
fields by changing the intermediate barrier thickness within the limits
of 2.5 - 7.5~nm. The application of the intermediate
barrier also appreciably affects the relative increase of
${I_{\sigma  + }}(B)$ enfeebling it. Herein
the barrier thickness is of minor importance (see inset in Fig.~\ref{10},
($a$)).

\section{POLARIZATION OF THE PL BANDS}

\begin{figure}\
 \centerline{\includegraphics [width=7cm]{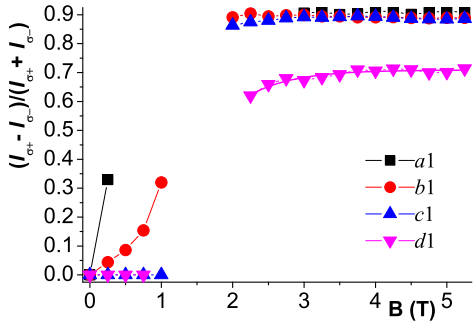}}
 \caption{Magnetic field dependence of the
${L_{4(2)}}$ band polarization.}
 \label{11}
\end{figure}

Polarization of the PL bands is traditionally determined as
$\rho  = ({I_{{\sigma ^ + }}} - {I_{{\sigma ^ -
}}})/({I_{{\sigma ^ + }}} + {I_{{\sigma ^ - }}})$. The
shortest $L_1^ - $ PL band disappears as
early as $B$ $>$~0.5~T
in any structure under study. Thus, the intermediate barrier does not
effect polarization of the ${L_1}$ band,
which arises as soon as a magnetic field is applied and reaches 100~\%
if $B$ $>$~0.5~T. The
state of matter is absolutely different for the longest
${L_4}$ PL band. Fig.~\ref{11} demonstrates it by
the example of the ${L_{4(2)}}$ bands.

As one can see, the ${L_{4(2)}}$ band is
not polarized up to $B$
$\approx{}$~1~T for the
$c1$ and $d1$
structures. At the same time, for the other two structures with the
thinnest intermediate barrier and without a barrier the
${L_{4(2)}}$ band starts to polarize as soon
as a magnetic field is applied. In the high magnetic fields, the
${L_{4(2)}}$ band polarization saturates.
The polarization of saturation is the largest in the
$a1$ structure without any intermediate
barrier (91~\%), somewhat smaller for the
$b1$ and $c1$
structures (89~-~88 \%), and noticeably
smaller for the $d1$ structure with the
thickest intermediate barrier (71~\%). The
polarization of the ${L_{4(1)}}$ band is the
same.

\section{DISCUSSION}
\label{discussion}

In our previous work \cite{ref13} we analyzed the PL spectra of the $a1$
structure and proposed the following interpretation of the origin of
different bands. \textit{Low magnetic fields}: (i) a transition between
the
$Zn_{0.9}Be_{0.05}Mn_{0.05}Se$ conduction band $E_{ZnBeMnSe}^C$ and the
energy level of an acceptor complex containing
$Mn$ $E_{ZnBeMnSe}^{Mn}$ forms the short wave
${L_1}$ band; (ii) a donor-acceptor
transition
$E_{ZnBeMnSe}^D$-$E_{ZnBeMnSe}^A$
in
$Zn_{0.9}Be_{0.05}Mn_{0.05}Se$ forms the ${L_2}$ band; (iii) an indirect
transition in real space between the $2D$
conduction band of the $ZnSe$ QW
$E_{ZnSe}^C$ and
$E_{ZnBeMnSe}^{Mn}$ forms the
${L_3}$ band. \textit{High magnetic fields}:
(i) a transition between $E_{ZnBeMnSe}^C$
and the barrier valence band
$E_{ZnBeMnSe}^V$forms the
${L_1}$ band; (ii) a transition
$E_{ZnBeMnSe}^D$-
$E_{ZnBeMnSe}^V$ forms the
${L_2}$ band; (iii) an indirect transition
$E_{ZnSe}^C$-
$E_{ZnBeMnSe}^V$ forms the
${L_3}$ band. The transitions within the
$ZnSe$ QW
$E_{ZnSe}^C$ - $E_{ZnSe}^V$ forms the
${L_4}$ band in any magnetic field. The
short wave component ${L_{4(1)}}$ of the
${L_4}$ band is formed by emission of the
light exciton, while the long wave component
${L_{4(2)}}$ is formed by emission of the
heavy exciton. There are no reasons to expect that the intermediate
$Zn_{0.943}Be_{0.057}Se$ barrier
changes the nature of the PL bands within both the
$ZnSe$ QW and the
$Zn_{0.9}Be_{0.05}Mn_{0.05}Se$
semimagnetic barrier. Therefore, the analysis that follows the origin
of the ${L_1}$,
${L_2}$, and
${L_4}$ bands is the same as in the other
three structures.

In order to separate the contribution of the barrier height and
exchange interaction in the field behavior of different PL bands, a
quantitative analysis is needed.

The ${L_1}$ band shift in a magnetic field
is caused by the giant Zeeman splitting and may be written as \cite{ref1, ref15,
ref16}:

\begin{equation}
E(B) = E(0) \mp (\chi {N_0})\tilde x < {S_Z} >
\label{eq1}
\end{equation}

\begin{equation}
\chi {N_0} = \alpha {N_0} - \beta {N_0}
\label{eq2}
\end{equation}

where $\tilde x$ is the effective
$Mn$ concentration,
$\alpha {N_0}$ and $\beta
{N_0}$ are the exchange integrals between the
$Mn$ ions and carriers for upper and lower
states responsible for this PL band, respectively, $ < {S_Z} > $ is the thermal average of the
$Mn$ spin given by:

\begin{equation} < {S_Z} >  = \frac{5}{2}{B_{{5 \mathord{\left/
{\vphantom {5 2}} \right.
\kern-\nulldelimiterspace} 2}}}\left[ {5{\mu _B}B/{k_B}(T +
{T_{eff}})} \right]\label{eq3}
\end{equation}

${B_{5/2}}$ is Brillouin function of the
argument in square brackets, ${\mu _B}$ is
the Bohr magneton, $k_{B}$ is the Boltzmann constant,
$T$ is the temperature, and
${T_{eff}}$ is an empirical parameter
representing antiferomagnetic interaction between the
$Mn$ ions. In our calculations, the
parameter ${T_{eff}}$ was taken to be equal
to 1.75~K in accordance with the empirical ratio obtained in
\cite{ref15} for $Zn_{1 - X}Mn_{X}Se$.

\begin{figure*}\
 \centerline{\includegraphics [width=13cm]{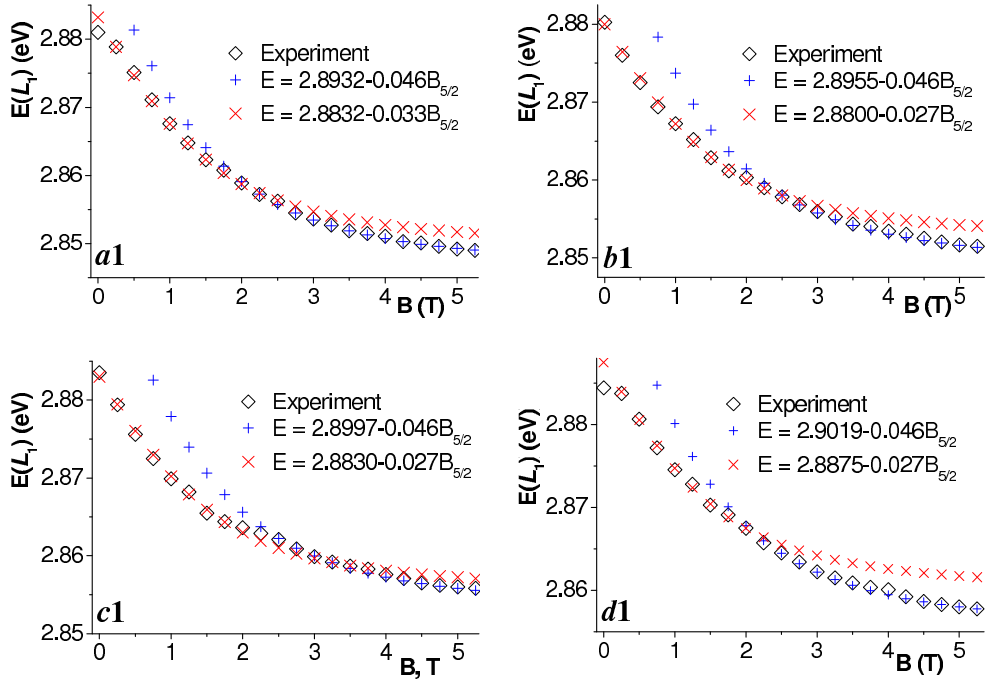}}
 \caption{Magnetic field dependence of the
E(${L_1}$) for the different structures
under study.}
 \label{12}
\end{figure*}

A comparison of the experimental and the calculated data for all
structures under study is presented in Fig.~\ref{12}. It clearly shows that
there are two ranges of magnetic fields for any structure where the
rates of the $L_1^ + $ band shift
($\chi {N_0}$) in the
$Zn_{0.9}Be_{0.05}Mn_{0.05}Se$ layer are different: they are smaller for low and larger for high
magnetic fields. We marked them by ($\chi {N_0}$)$_{1L}$ and ($\chi {N_0}$)$_{1H}$, respectively. The parameter
$E(0)$ is also different for the ranges of
low and high magnetic fields. We marked them
${E_{1L}}(0)$ and
${E_{1H}}(0)$, respectively.

One can see in Fig.~\ref{12} that many of the parameters of the ${L_1}$ band in the $Zn_{0.9}Be_{0.05}Mn_{0.05}Se$ layer are different not only for low and high magnetic fields. They are also different for the structures with and without the intermediate $Zn_{0.943}Be_{0.057}Se$ barrier as well as for the structures of different barrier thickness. Let us examine these parameters.

\begin{figure}\
 \centerline{\includegraphics [width=7cm]{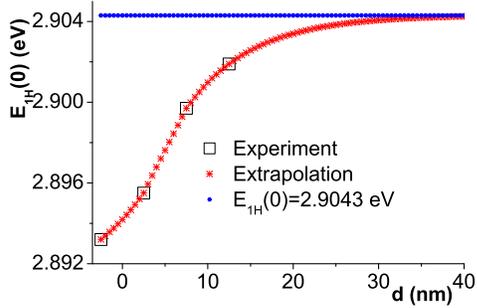}}
 \caption{${E_{1H}}(0)$ vs $d$ of the layers contacting with the $Zn_{0.9}Be_{0.05}Mn_{0.05}Se$ semimagnetic barrier in the investigated structures. $d$ is positive for the $Zn_{0.943}Be_{0.057}Se$ layers (compression of $Zn_{0.9}Be_{0.05}Mn_{0.05}Se$ layer) and negative for $ZnSe$ layer (tension of $Zn_{0.9}Be_{0.05}Mn_{0.05}Se$ layer).}
 \label{13}
\end{figure}

${E_{1H}}(0)$ has a sense of a band gap of the semimagnetic $Zn_{0.9}Be_{0.05}Mn_{0.05}Se$ barrier in zero magnetic fields. This parameter increases monotonously from 2.8932~eV for the barrier in the $a1$ structure to 2.9019~eV for the same barrier in the $d1$ structure (Fig.~\ref{12}). We explain the observed changes of the parameters ${E_{1H}}(0)$ as well as of the energy of ${L_1}$ bands by both: (i) the effect of the strains of the $Zn_{0.9}Be_{0.05}Mn_{0.05}Se$ barrier surface layers caused by the action of the adjacent $ZnSe$ or $Zn_{0.943}Be_{0.057}Se$ layers on the band gap value; (ii) the dominant contribution of the strained surface layers in forming the emission bands of the semimagnetic barrier. It is well known that in the absence of strain, the maxima of the heavy- and light-hole valence bands are degenerate in the zinc-blende semiconductors. The biaxial strain shifts and splits the heavy- and light-hole bands. If the strain is compressive, the band gap increases and coincides with the heavy-hole-derived band gap. In the case of biaxial tension, the band gap shrinks and is associated with the light-hole transition \cite{ref17}. The layers of the structures under study have different lattice constants:
$a$($ZnSe$)~{\large=~}5.6684~\AA{},
$a$($Zn_{0.943}Be_{0.057}Se$)~=~5.6382~\AA{},
$a$($Zn_{0.9}Be_{0.05}Mn_{0.05}Se$)~=~5.6592 \AA{} \cite{ref18,ref19}. Therefore, they deform each other:
$ZnSe$ tenses the surface layers of
$Zn_{0.9}Be_{0.05}Mn_{0.05}Se$, and
$Zn_{0.943}Be_{0.057}Se$ compresses
them. Thus, the
$Zn_{0.9}Be_{0.05}Mn_{0.05}Se$ surface layers that contact with the $ZnSe$
layer, have a smaller band gap, and contacting with the
$Zn_{0.943}Be_{0.057}Se$ layer have a
larger band gap than the strainless one. It enables us to construct a
dependence of the band gap of the strain
$Zn_{0.9}Be_{0.05}Mn_{0.05}Se$ layers on the thickness of both the contact
$ZnSe$ and
$Zn_{0.943}Be_{0.057}Se$ layers which
cause these strains. This dependence is presented in Fig.~\ref{13}.

Fig.~\ref{13} shows that the band gap of the
$Zn_{0.9}Be_{0.05}Mn_{0.05}Se$ strain
layers asymptotically approaches the value 2.9043~eV under compression
by the $Zn_{0.943}Be_{0.057}Se$ layer
when $d(Zn_{0.943}Be_{0.057}Se)$
approaches 30~nm. Extrapolation of the data in Fig.~\ref{13} also shows that
the band gap of the strainless
$Zn_{0.9}Be_{0.05}Mn_{0.05}Se$ is
equal to 2.894~eV. Thus, the biaxial strain of the
$Zn_{0.9}Be_{0.05}Mn_{0.05}Se$ layer
by the $Zn_{0.943}Be_{0.057}Se$ layer
can increase the band gap of the former by approximately 10~meV.

The obtained dependences of
$E(L_1^ + )$ from
$B$ may be explained by the dependence of
the band gap of
$Zn_{0.9}Be_{0.05}Mn_{0.05}Se$ on the
strains if the strained contact layers provide the main contribution to
the emission transitions. Note that the importance of the strained
heterointerface for localization of excitons was already emphasized in initial investigations of the II-VI strained-layer heterostructures\cite{ref20}.

The value ${E_{1H}}(0) - {E_{1L}}(0)$ has a
sense of an acceptor level depth of the
$Mn$ complex in the
$Zn_{0.9}Be_{0.05}Mn_{0.05}Se$
strained surface layers. It is approximately one and a half times
smaller in the compressed layers in comparison with the tensed one: 10~meV for the
$Zn_{0.9}Be_{0.05}Mn_{0.05}Se$ barrier
in the $a1$ structure against
(15.5~$\pm{}$~1.2) meV for the other three
structures.

The values of ${(\chi {N_0})_{1H}}$ and
${(\chi {N_0})_{1L}}$ are determined by the
exchange interaction splitting the electronic band and impurity levels.
It is obvious from Fig.~\ref{12} that the ${(\chi{N_0})_{1H}}$ values are the same for all structures. It means
that the $\alpha {N_0}$ and
$\beta {N_0}$ exchange integrals for
$C$ and $V$ bands of
$Zn_{0.9}Be_{0.05}Mn_{0.05}Se$ layer
develop with no effect of the $2D$ free
carriers of the $ZnSe$ QW. On the contrary,
the ${(\chi {N_0})_{1L}}$ value is the same
only for the structures with the intermediate
$Zn_{0.943}Be_{0.057}Se$ barrier. For
the $a1$ structure without any intermediate
barrier it is larger by approximately 22~\%. It means that the exchange
integral for electronic states of the
$Mn$ complex in the structures
under study develops appreciably with participation of the
$2D$ free carriers of the
$ZnSe$ QW. In accordance with our previous
analysis \cite{ref13} the band exchange integrals for  the
$Zn_{0.9}Be_{0.05}Mn_{0.05}Se$ layer
are equal to $\alpha {N_0}$~=~0.104~eV,
$\beta {N_0}$~=~-~0.264~eV. The
value of the exchange integral for electronic states of the
$Mn$ complex in the
$a1$ structure is equal to 0.156 eV. Using
the obtained ${(\chi {N_0})_{1L}}$ value for
the $b1$, $c1$,
and $d1$ structures we see that the
application of intermediate
$Zn_{0.943}Be_{0.057}Se$ barrier
decreases this value to 0.112 eV, i.e., by approximately 40~\%.

We explain these results in the following way. There are two types of
free carriers contributing to the exchange interaction in the strained
surface layers of the
$Zn_{0.9}Be_{0.05}Mn_{0.05}Se$
barrier: $3D$ carriers of the
$Zn_{0.9}Be_{0.05}Mn_{0.05}Se$ barrier
and $2D$ carriers of the
$ZnSe$ QW. The wave functions of
$3D$ carriers are quite extended and span a
large number of lattice sites of the
$Zn_{0.9}Be_{0.05}Mn_{0.05}Se$
barrier. On the contrary, only the tails of the wave functions of
$2D$ carriers penetrate from QW into the
barrier. This penetration decreases or is absent when the intermediate
$Zn_{0.943}Be_{0.057}Se$ barrier appears
between the $ZnSe$ and
$Zn_{0.9}Be_{0.05}Mn_{0.05}Se$ layers.
Thus, we can unambiguously conclude that
$2D$ carriers of the
$ZnSe$ QW give negligible contribution to
the formation of exchange integrals for $C$ and
$V$ bands of the
$Zn_{0.9}Be_{0.05}Mn_{0.05}Se$
strained contact layers but their contribution to the exchange
interaction between the $Mn$ complexes in these layers is can be described by mixing of the
$3{d^5}$ levels of Manganese with the states
of 2$D$ free carriers of the $ZnSe$ QW. This
is possible only if the $Mn$ complex concentration in the strained layers is larger than the
concentration of $Mn$ in the sites
of crystal lattice. In other words, most probably the
$Mn$ complexes in
$Zn_{0.9}Be_{0.05}Mn_{0.05}Se$ develop
into the strained layers of a heterocontact.

Let us now consider the long wave ${L_4}$
PL band. The layer of the $ZnSe$ QW
experiences compression on the part of both
$Zn_{0.9}Be_{0.05}Mn_{0.05}Se$ and
$Zn_{0.943}Be_{0.057}Se$ layers. As a
result, its heavy- and light-hole bands split and the heavy-hole band
defines the band gap in any structures under study. Therefore, the
heavy excitons $\left| { \pm 1} \right\rangle  = \left| { \mp 1/2, \pm 3/2} \right\rangle $ form the long wave
component $L_{4(2)}^ \pm $ of
${\sigma ^ + }$ and
${\sigma ^ - }$-polarization, respectively,
and the light excitons $\left| { \pm 1} \right\rangle
 = \left| { \pm 1/2, \pm 1/2} \right\rangle $ form the short
wave component $L_{4(1)}^ \pm $ of the
${L_4}$ band.

\begin{figure}\
 \centerline{\includegraphics [width=7cm]{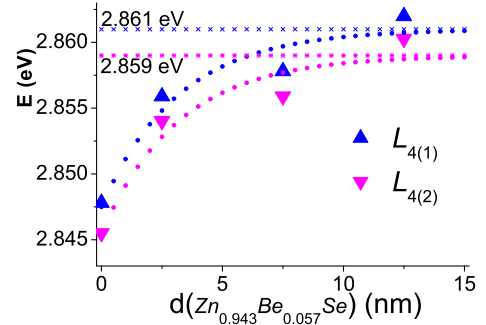}}
 \caption{Energy of the $L_{4(1)}^ \pm $ and
$L_{4(2)}^ \pm $ bands vs. the intermediate
$Zn_{0.943}Be_{0.057}Se$ barrier
thickness in zero magnetic fields and their approximations.}
 \label{14}
\end{figure}

A splitting between the energy of heavy and light excitons in zero
magnetic fields is equal to (2.0~$\pm{}$~0.3)~meV
for all structures. At the same time, the energies of both heavy and
light excitons are different for different structures and increase with
an increased thickness of the intermediate nonmagnetic barrier
$Zn_{0.943}Be_{0.057}Se$ between the
$Zn_{0.9}Be_{0.05}Mn_{0.05}Se$ and $ZnSe$ layers. The dependences of
$E$($L_{4(1)}^ \pm $)
and $E$($L_{4(2)}^ \pm
$) on the thickness of the intermediate barrier in case of the
absence of magnetic field are shown in Fig.~\ref{14}.

One can see that both
$E$($L_{4(1)}^ \pm $)
and $E$($L_{4(2)}^ \pm
$) increase if
$d$($Zn_{0.943}Be_{0.057}Se$)
increases and asymptotically approach the energy 2.861 and 2.859 eV,
respectively. The cause of these changes is the same as in the case of
the ${L_1}$ bands: the deformation effects.
There is the following ratio between the lattice constants of the
layers of the structures:
$a(ZnSe)$
$>$
$a(Zn_{0.9}Be_{0.05}Mn_{0.05}Se)$
$>$
$a(Zn_{0.943}Be_{0.057}Se)$.
As a result, the
$Zn_{0.943}Be_{0.057}Se$ layer more
strongly compresses the $ZnSe$ QW layer than
the $Zn_{0.9}Be_{0.05}Mn_{0.05}Se$
layer and its deformation effect increases if
$d$($Zn_{0.943}Be_{0.057}Se$)
increases. Therefore, the $ZnSe$ layer band
gap increases and the exciton energy increases too. For the
$a1$ structure in zero magnetic fields
$E$($L_{4(1)}^ \pm $)
=~2.8478,~and  $E$($L_{4(2)}^ \pm
$) =~2.8455 eV. It is seen that under the effect of the
intermediate barrier, the heavy-hole-derived band gap of the
$ZnSe$ QW increases by 13 meV extra.

Application of a magnetic field splits the band edges in the well.
However, the Zeeman splitting for nonmagnetic
$ZnSe$ should be negligible taking into
account the g-factor value for electrons and holes\cite{refLdB99}. Actually
$E$($L_{4(1)}^ \pm $)
and $E$($L_{4(2)}^ \pm
$) negligibly depends on $B$ only
for the $b1$,
$c1$, and $d1$
structures and only in magnetic fields $B$
$<$ (2.0 $\div{}$ 2.5)~T
(Fig.~\ref{5}). For higher magnetic fields, the energy of
${L_4}$ exciton bands appreciably decreases.
There are more essential changes in the energy positions of both
$L_{4(1)}^ \pm $ and
$L_{4(2)}^ \pm $ bands in a magnetic field
for the $a1$ structure, as we have
emphasized above. It is obvious that the exchange interaction between
free carriers of the QW and $Mn$ ions of the semimagnetic barrier, which is further modified by the
intermediate nonmagnetic barrier, causes the observed changes.

\begin{figure}\
 \centerline{\includegraphics [width=7cm]{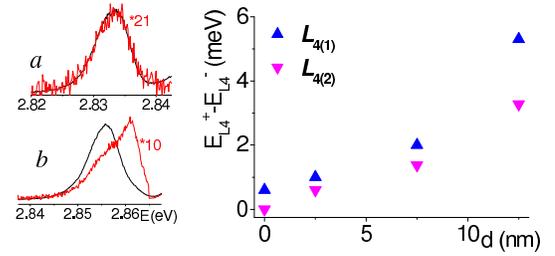}}
 \caption{The experimental PL spectra of the
$a1$ ($a$) and
$d1$ structures ($b$) of
${\sigma ^ + }$ (black lines) and
${\sigma ^ - }$-polarization (red lines) for
$B$ =~5.25~T in the energy range of the
${L_4}$ bands. Right - the shift between the energy
positions of $L_4^ + $ and
$L_4^ - $ bands vs.
$d$($Zn_{0.943}Be_{0.057}Se$) at 5.25~T.}
 \label{15}
\end{figure}

The $a1$ structure is the structure with a
shallow nonmagnetic QW having the adjacent layer of semimagnetic
semiconductor. Therefore, the effective depth of the well, given by the
difference between the positions of the band edges in the adjacent
layers, is strongly dependent on the magnetic field, and in a shallow
well the energy of the size quantization levels is strongly dependent
on the well depth. For the other structures, the situation is radically
different. Herein the positions of the band edges of the
$Zn_{0.943}Be_{0.057}Se$ layers adjacent
to the QW negligibly depend on $B$ and the
height of the intermediate barrier also negligibly depends on
$B$. At the same time, an effective barrier height
for electrons and holes appreciably depends on
$B$ because the position of the band edges
of the $Zn_{0.9}Be_{0.05}Mn_{0.05}Se$
layer depends on $B$. It changes a range of
penetration of the wave functions of the
$ZnSe$ QW well free carriers in the barrier
($Zn_{0.9}Be_{0.05}Mn_{0.05}Se$ +
$Zn_{0.943}Be_{0.057}Se$) because the
boundary conditions on the QW edge
($Zn_{0.943}Be_{0.057}Se$~/ $ZnSe$)
change. However, a change of the height of the intermediate and effective
barriers has got a different effect on the shift of both
$L_{4(1)}^ \pm $ and
$L_{4(2)}^ \pm $ bands. As soon as an
intermediate barrier height changes, the
${L_4}$ band position also changes. In case
of an effective barrier, the situation is different. As it follows from
the obtained data, the $E_{ZnBeMnSe}^{{V^ +
}}$ edge should go down appreciably lower than the
$E_{ZnSe}^{{V^ + }}$ edge before the
${L_4}$ bands change their position. In
accordance with the energy diagram of the
$a1$ structure \cite{ref13}, the offset between the $C$ and
$V$ bands of the
$Zn_{0.9}Be_{0.05}Mn_{0.05}Se$ and
$2D$ $ZnSe$
layers distributed as
$\Delta{}$E$_{C}$/$\Delta{}$E$_{V}$~=~60/40
which is typical of these materials \cite{ref7}. The intermediate barrier somewhat decreases the value
$({E_{1H}}(0) - {E_{L4(2)}}(0))$, which determines the mutual location of the $Zn_{0.9}Be_{0.05}Mn_{0.05}Se$ and
$2D$ $ZnSe$ layer band edges on the energy scale. Its magnitude for the $a1$ structure is equal to 47.7~meV, and the averaged magnitude for the other three structures is equal to (42.3~$\pm{}$~1.5) meV. If we share this value in proportion $\Delta {E_C}/\Delta {E_V} = 60:40$~we find that $E_{ZnSe}^V$-$E_{ZnBeMnSe}^V$ in zero magnetic field for the $b1$, $c1$, and $d1$ structures is equal to 16.9 meV. The energy of $L_{4(1)}^ \pm $ and $L_{4(2)}^ \pm $ bands start to decrease only if the $E_{ZnBeMnSe}^{{V^ + }}$ goes down to approximately 24.5~meV ($B \approx{}$2~T). However, one needs to remember that the effective barrier for both electrons and holes
forming the $L_4^ + $ bands goes down when $B$ increases. On the contrary, the
effective barrier for electrons and holes forming the
$L_4^ - $ bands goes up. At the same time,
the energy of $L_{4(1)}^ \pm $ and
$L_{4(2)}^ \pm $ pair decreases in the range
of high magnetic fields. It entirely bears a resemblance to the
behavior of both $L_{4(1)}^ \pm $ and
$L_{4(2)}^ \pm $ bands in the
$a1$ structure in this magnetic range. In
our previous work \cite{ref13} we explained such a behavior of different polarized exciton bands in
the QW by the influence of the spin-flip processes caused by the
degeneration of energy levels of the
$ZnSe$ QW and the
$Zn_{0.9}Be_{0.05}Mn_{0.05}Se$ layer.
We believe that the same processes also occur in the structures with
the intermediate nonmagnetic barrier but the barrier presence somewhat
changes the situation. The barrier weakens the interaction between the
$ZnSe$ QW and the
$Zn_{0.9}Be_{0.05}Mn_{0.05}Se$ layer.
As a result, the spin-flip processes also weaken and the differently
polarized ${L_4}$ bands gradually drift
apart. The data in Fig.~\ref{15} clearly show this.

Let us examine the ${L_2}$ band. It is
formed by the emission transition inside the semimagnetic layer
$Zn_{0.9}Be_{0.05}Mn_{0.05}Se$ but the
presence of the intermediate barrier in the structure and its thickness
appreciably changes this band behavior. First of all, this barrier
changes the energy positions of the donor
$E_{ZnBeMnSe}^D$ and acceptor
$E_{ZnBeMnSe}^A$ levels in the
$Zn_{0.9}Be_{0.05}Mn_{0.05}Se$ surface
strain layers in zero magnetic field, which form the
${L_2}$ band. A sum of
$E_{ZnBeMnSe}^D$ and
$E_{ZnBeMnSe}^A$ may be determined as the
difference between ${E_{1H}}(0)$ 
and ${E_{L2}}(0T)$. This difference is equal
to 31~meV for the $a1$ structure and is
larger by approximately 3 meV for the other three structures. Fig.~\ref{5}
data make it possible to divide a contribution of both
$E_{ZnBeMnSe}^D$ and
$E_{ZnBeMnSe}^A$ shifts under conversion
from the tensed to compressed contact layers of
$Zn_{0.9}Be_{0.05}Mn_{0.05}Se$ in
formation of the mentioned difference. In the
$b1$, $c1$, and
$d1$ structures, both
$L_2^ + $ and $L_1^ +
$ bands intersect in a magnetic field approximately over
1.9~T. In accordance with the rate of both
$E_{ZnBeMnSe}^{{C^ + }}$ and
$E_{ZnBeMnSe}^{{V^ + }}$ level shift, it
corresponds to
$E_{ZnBeMnSe}^D$~$\approx{}$~10
meV and
$E_{ZnBeMnSe}^A$~$\approx{}$~24
meV. If we compare these values with the same ones for the
$a1$ structure \cite{ref13}, we see that $E_{ZnBeMnSe}^D$ decreases
approximately by 5.5 meV and
$E_{ZnBeMnSe}^A$ increases approximately by
8.5 meV under conversion from a tension deformation to a compression
deformation of the
$Zn_{0.9}Be_{0.05}Mn_{0.05}Se$ layers.
We explain this in the following way. Under deformation, both spectrum
and wave functions of the electrons for the degenerate band are
determined by the solution of the wave equation with Hamiltonian having
an addition, which determines of the band splitting. This splitting
leads to a reconstruction of the impurity spectrum especially if the
splitting of the degenerated band grows up to the energy of impurity
ionization $E$$_{i}$ \cite{ref21}. In our case, it corresponds to an acceptor center case. The case for
a donor center is different. A conduction band of
$Zn_{0.9}Be_{0.05}Mn_{0.05}Se$ is
non-degenerate. For a non-degenerate band, the change of energy of an
impurity center
$\Delta{}$$E$$_{i}$/$E$$_{i}$
is only related to the change of the free carrier effective mass and is
approximately
$\Delta{}$$m$/$m$
\cite{ref21}. For estimation, $E$i may be equated to the
average value of $E_{ZnBeMnSe}^D$ for the
structures with and without the intermediate barrier while
$\Delta{}$$E$$_{i}$ may be
equated to the deviation from this average value. Then, we conclude that
the deformations originated by the lattice mismatches can cause the
change of the free electron effective mass\cite{mass} by about 20~\% in the
structures under study.

Another important feature of the ${L_2}$
band is its comparative shift relatively to the $L_1^+ $ band if a magnetic field is applied. The
$L_2^ - $ band does not change its energy
position either before or after intersecting with the
$L_1^ + $ band position. This means that the
spin-up states of impurity electrons do not mix with the spin-down band
electron states and form the resonance state in both
$C$ and $V$ bands of
$Zn_{0.9}Be_{0.05}Mn_{0.05}Se$ in high
magnetic fields. The spin-down states of impurity electrons mix with
the spin-down band electron states in high magnetic fields. Therefore,
the energy of the $L_2^ + $ band decreases
if the $E_{ZnBeMnSe}^A$ level intersects the
$E_{ZnBeMnSe}^{{V^ + }}$ edge (the
$a1$ structure, Fig.~\ref{5}) or the band
disappears if both $E_{ZnBeMnSe}^A$ and
$E_{ZnBeMnSe}^D$ levels intersect
$E_{ZnBeMnSe}^{{V^ + }}$ and
$E_{ZnBeMnSe}^{{C^ + }}$ edges, respectively
(both $b1$ and
$c1$ structure, Fig~\ref{5}). It is not entirely
clear why the resonance spin-down states of impurity electrons appear
in the $C$ and $V$ band of
strained
$Zn_{0.9}Be_{0.05}Mn_{0.05}Se$ layers
in the $d1$ structure with the thickest
intermediate barrier in high magnetic field as well as why the spin-up
states of impurity electrons disappear in both
$a1$ and $b1$
structures in relatively low magnetic fields.

Let us now examine some aspects of the PL intensity related to the
presence of the $Zn_{0.943}Be_{0.057}Se$
intermediate barrier in the structures under study and its effect on
the transfer of magnetic interaction between the semimagnetic
$Zn_{0.9}Be_{0.05}Mn_{0.05}Se$ layer
and nonmagnetic $ZnSe$ QW. The first one is
the barrier effect on the intensity of the
$2D$ excitons of the
$ZnSe$ QW in zero magnetic fields. As one
can see in Fig.~\ref{8}, this intensity increases if the intermediate barrier
is applied and
$d$($Zn_{0.943}Be_{0.057}Se$)
increases. We explain this in the following way. Applying the
intermediate barrier we equalize the strain on the both sides of the
QW. The strain equalization becomes more effective when the
intermediate barrier thickness increases. A decrease of the structure
inhomogeneity naturally leads to an increase of emitting recombination.

The second aspect is the barrier effect on the field dependence of the
intensity of ${L_4}$ bands in low
magnetic fields. Giant Zeeman splitting of the band edges of the
$Zn_{0.9}Be_{0.05}Mn_{0.05}Se$ layer
is an immediate cause of this effect. Applying a magnetic field we
decrease both $E_{ZnBeMnSe}^{{C^ + }}$ and
$E_{ZnBeMnSe}^{{V^ + }}$ edges of the
$C$ and $V$ bands of the
$Zn_{0.9}Be_{0.05}Mn_{0.05}Se$ layer.
As a result, the
$Zn_{0.943}Be_{0.057}Se$ barrier
confines the thermalized carriers in this layer and counteracts their
transfer into the $ZnSe$ QW layer. The
larger is magnetic field, the stronger is the carrier confinement in
the $Zn_{0.9}Be_{0.05}Mn_{0.05}Se$
layer. Since the PL intensity is lower if a concentration of the
recombining electrons and holes is smaller, both
$I$($L_{4(1)}^ + $) and
$I$($L_{4(2)}^ + $)
decrease if $B$ increases.

A recombining carrier concentration is not only a factor defined PL
intensity. It especially depends on the emission probability, which,
for its part, is proportional to a density of states of the free carriers
\cite{ref22}. A magnetic field applied transversely to the structure layers
transforms the free $2D$ carriers in the
$ZnSe$ QW to 0$D
$carrier if $B$ increases to the
quantum strong limit. In this field range a density of states of the
carrier is defined by $B$ and increases if
$B$ increases. Therefore, the intensity of
$L_4^ + $ bands has to change a decrease for
an increase if a magnetic field passes in the range of quantum magnetic
fields. As one can see from Fig.~\ref{9} (right),
$I$($L_4^ + $) starts
increasing at $B>$~0.75~T. Using
$m$$_{hh}$~=~0.6~$m$$_{o}$
\cite{ref23}, for this $B$ value we obtain a hole
cyclotron energy $\hbar {\omega_c}$~$\approx{}$~0.14~meV. The thermal
energy of the experiment is
$k$$_{o}$T~$\approx{}$~0.1~meV.
Therefore, for
$B$~$>$~0.75~T, the
condition of quantum magnetic fields is already in progress. As a
result, two factors determine the further behavior of
$I$($L_4^ + $) in a
magnetic field: a confinement of the free carrier in the
$Zn_{0.9}Be_{0.05}Mn_{0.05}Se$ layer
by the $Zn_{0.943}Be_{0.057}Se$ barrier
and an increase of the state density of the 0$D$
carriers in the $ZnSe$ QW if
$B$ increases. The latter factor dominates
and $I$($L_4^ + $)
increases.

\section{CONCLUSIONS}

In this paper we have reported the measurements of luminescence of the 150~nm $Zn_{0.9}Be_{0.05}Mn_{0.05}Se$~/ $d$~nm~$Zn_{0.943}Be_{0.057}Se$~/ 2.5~nm~$ZnSe$~/ 30~nm~$Zn_{0.943}Be_{0.057}Se$ structures as a function of thickness $d$
of the intermediate nonmagnetic barrier
$Zn_{0.943}Be_{0.057}Se$ between
the
$Zn_{0.9}Be_{0.05}Mn_{0.05}Se$ semimagnetic
barrier and $ZnSe$ QW and magnetic field at
the low temperature 1.2~K. Strong evidence has been obtained that the
intermediate nonmagnetic barrier: (i) changes the energy of the PL
bands in both the $ZnSe$ QW and
$Zn_{0.9}Be_{0.05}Mn_{0.05}Se$ semimagnetic
barrier layers; (ii) increases the total PL intensity of the
structures; (iii) decreases the degree of circular polarization of the
QW exciton emission in the structure; (iv) extinguishes the PL band
caused by the indirect transitions in real space between the
$2D$ conduction band of the 
$ZnSe$ QW and both the
$Mn$ complex and the valence band
of the
$Zn_{0.9}Be_{0.05}Mn_{0.05}Se$~layer.
The obtained data enable us to conclude that the emission bands
appearing in the semimagnetic
$Zn_{0.9}Be_{0.05}Mn_{0.05}Se$~barrier
of the structures under study are formed in the contact layers strained
by the intermediate
$Zn_{0.943}Be_{0.057}Se$~or
$ZnSe$ layers. The shifts of the band gap as
well as of the donor and acceptor levels under the effect of biaxial
compression of the
$Zn_{0.9}Be_{0.05}Mn_{0.05}Se$ layer
by the $Zn_{0.943}Be_{0.057}Se$ layer
are estimated.

It is revealed that there are two different rates
($\chi {N_0}$)$_{1H}$ and
($\chi {N_0}$)$_{1L}$ of the shift of the
short wave bands of the PL spectra of the structure under study in a
magnetic field caused by giant Zeeman splitting. The larger rate
($\chi {N_0}$)$_{1H}$ is observed in the
high magnetic fields and corresponds to the emission transition between
$C$ and $V$ bands of the
$Zn_{0.9}Be_{0.05}Mn_{0.05}Se$~barrier.
A smaller rate ($\chi {N_0}$)$_{1L}$ is
observed in low magnetic fields and corresponds to the emission
transition between the
$Zn_{0.9}Be_{0.05}Mn_{0.05}Se$~conduction
band and an energy level of the acceptor complex containing
$Mn$. The intermediate nonmagnetic barrier
$Zn_{0.943}Be_{0.057}Se$ does not change
the ($\chi {N_0}$)$_{1H }$value and
accordingly its constituents $\alpha {N_0}$
and $\beta {N_0}$ values, which are equal to
$\alpha{N_0}$~=~0.104~eV and~$\beta{N_0}$~=~-0.264~eV. At the same time, it
decreases the ($\chi {N_0}$)$_{1L}$ value by
approximately 22~\% which we interpret as a decrease of the exchange
integral for electronic states of the
$Mn$ complex in the
$Zn_{0.9}Be_{0.05}Mn_{0.05}Se$~barrier
by approximately 40~\% (from 0.156 to 0.112 eV) under the effect of the
intermediate $Zn_{0.943}Be_{0.057}Se$ barrier. This supports the assumption that: (i) the deformation of the
$Zn_{0.9}Be_{0.05}Mn_{0.05}Se$~layers
plays a key role in forming the $Mn$ complexes; (ii) the $2D$ carriers of the
$ZnSe$ QW provide a substantial contribution
to the formation of the exchange integral for the
$Mn$ complexes in strained layers.
The $Zn_{0.943}Be_{0.057}Se$ intermediate
barrier changes the effect of giant Zeeman splitting of the
semimagnetic
$Zn_{0.9}Be_{0.05}Mn_{0.05}Se$ barrier
energy levels on a move of the energy levels of
$ZnSe$ QW in a magnetic field and a
polarization of the QW exciton emission.

\section*{acknowledgments}
AS would like to thank the Concept for the Future in the Excellence Initiative at KIT for financial support.

\bibliographystyle{unsrt}

\begin{thebibliography}{00}
 \bibitem{ref1} J.K. Furdyna, J. Appl. Phys. \textbf{64}, R29 - R64 (1988)
 \bibitem{ref2} Semiconductor Spintronics and Quantum Computation, edited by D.D. Awschalom, D. Loss, and N. Samarth (Springer-Verlag, Berlin, 2002).
 \bibitem{ref3} Spin Physics in Semiconductors, edited by M.I. Dyakonov (Springer-Verlag, Berlin, 2008)
 \bibitem{ref4} B.T. Jonker, Y.D. Park, B.R. Bennett, H.D. Cheong, G. Kioseoglou, A. Petrou, Phys. Rev. B \textbf{62}, 8180 (2000).
 \bibitem{ref5} C. Gould, A. Slobodskyy, T. Slobodskyy, P. Grabs, C.R. Becker, G. Schmidt, and L.W. Molenkamp, physica status solidi (b) \textbf{241}, 700 (2004).
 \bibitem{ref6} A.A. Maksimov, D.R. Yakovlev, J. Debus, I.I. Tartakovskii, A. Waag, G. Karczewski, T. Wojtowicz, J. Kossut, and M. Bayer, Phys. Rev. B \textbf{82}, 035211 (2010).
 \bibitem{ref7} M. Kim, C.S. Kim, S. Lee, J.K. Furdyna, and M. Dobrowolska, J. Cryst. Growth, \textbf{214/215}, 325-329 (2000).
 \bibitem{ref8} A. Slobodskyy, C. Gould, T. Slobodskyy, C.R. Becker, G. Schmidt, and L.W. Molenkamp, Phys. Rev. Lett. \textbf{90}, 246601 (2003).
 \bibitem{ref9} M.K. Kneip, D.R. Yakovlev, M. Bayer, T. Slobodskyy, G. Schmidt, and L.W. Molenkamp, Appl. Phys. Lett. \textbf{88}, 212105 (2006).
 \bibitem{ref10} G. V. Astakhov, R. I. Dzhioev, K. V. Kavokin, V. L. Korenev, M. V. Lazarev, M. N. Tkachuk, Yu. G. Kusrayev, T. Kiessling, W. Ossau, and L. W. Molenkamp, Phys. Rev. Lett. \textbf{101}, 076602 (2008).
 \bibitem{ref11} G. V. Astakhov, M. M. Glazov, D. R. Yakovlev, E. A. Zhukov, W. Ossau, L. W. Molenkamp, and M. Bayer, Semicond. Sci. Technol. \textbf{23}, 114001 (2008).
 \bibitem{ref12} M.I. Dyakonov and V.I. Perel, \textquotedblright{}Optical orientation in a system of electrons and lattice nuclei in semiconductors. Theory\textquotedblright{}, Sov. Phys. JETP \textbf{38}, 177 (1974).
 \bibitem{ref13} D.M. Zayachuk, T. Slobodskyy, G. Astakhov, C. Gould, G. Schmidt, W. Ossau, and L.W. Molenkamp, EPL \textbf{91}, 67007 (2010).
 \bibitem{ref14} D. Keller, D.R. Yakovlev, B. K\"{o}nig, W. Ossau, Th. Gruber, A. Waag, L.W. Molenkamp, and A.V. Scherbakov, Phys. Rew. B \textbf{65}, 035313 (2001).
 \bibitem{ref15} H. Hoffmann, G. V. Astakhov, T. Kiessling, W. Ossau, G. Karczewski, T. Wojtowicz, J. Kossut, and L. W. Molenkamp, Phys. Rev. B \textbf{74}, 073407 (2006).
 \bibitem{ref16} N. Dai, L.R. Ram-Mohan, H. Luo, G.L. Yang, F.C. Zhang, M. Dobrowolska, and J.K. Furdyna, Phys. Rew. B \textbf{50}, 18153 (1994).
 \bibitem{ref17} B. Rockwell, H. R. Chandrasekhar, M. Chandrasekhar, A. K. Ramdas, M. Kobayashi, and R. L. Gunshor, Phys. Rev. B \textbf{44}, 11307 (1991).
 \bibitem{ref18} W. Faschinger, M. Ehinger, T. Schallenberg, M. Korn, Appl. Phys. Lett. 74, 3404 (1999).
 \bibitem{ref19} A.R. Denton and N.W. Ashcroft, Phys. Rev. A \textbf{43}, 3161 (1991).
 \bibitem{ref20} X.-C. Zhang, S.-K. Chang, A.V. Nurmikko, L.A. Kolodziejski, R.L. Gunshor, and S. Datta, Phys. Rev. B \textbf{31}, 4056 (1985).
 \bibitem{refLdB99} Landolt-B\"{o}rnstein. II-VI and I-VI Compounds; Semimagnetic Compounds, volume III, chapter 41b. (Springer, Berlin, 1999).
 \bibitem{ref21} G.L. Bir and G.E. Picus, Symmetry and Deformation in Semiconductors (Science, Moscow, 1972) (in Russian).
 \bibitem{mass} Charles Kittel Introduction to Solid State Physics (7th Edition ed.) (Wiley, New York, 1996).
 \bibitem{ref22} Peter Y. Yu Manuel Cardona, Fundamentals of Semiconductor (Springer, Berlin, 2005).
 \bibitem{ref23} H.J. Lozykowsky, V.K. Shastri, J. Appl. Phys. \textbf{69}, 3235 (1991).
\end{thebibliography}

\end{document}